\documentclass[utf-8]{article}
\usepackage{url,hyperref,lineno,microtype,bm,graphicx,authblk,amsmath}
\usepackage[left=2cm, right=2cm, top=1.785cm, bottom=2.0cm]{geometry}
\usepackage[onehalfspacing]{setspace}
\renewcommand{\d}{\text{d}}

\author[a,*]{Olav Galteland}
\author[a]{Michael T. Rauter}
\author[a]{Kevin K. Varughese}
\author[a]{Dick Bedeaux}
\author[a]{Signe Kjelstrup}
\affil[a]{PoreLab, Department of Chemistry, Norwegian University of Science and Technology}
\affil[*]{Corresponding author: olav.galteland@ntnu.no}

\title{Defining the pressures of a fluid in a nanoporous, heterogeneous medium} 
\begin{document}
\maketitle
\begin{abstract}
We describe the thermodynamic state of a single-phase fluid confined to a porous medium with Hill's thermodynamics of small systems, also known as nanothermodynamics. 
This way of defining small system thermodynamics, with a separate set of control variables, may be useful for the study of transport in non-deformable porous media, where presently no consensus exists on pressure computations. For a confined fluid, we observe that there are two pressures, the integral and the differential pressures. We use molecular simulations to investigate and confirm the nanothermodynamic relations for a representative elementary volume (REV). For a model system of a single-phase fluid in a face-centered cubic lattice of solid spheres of varying porosity, we calculate the fluid density, fluid-solid surface tension, replica energy, integral pressure, entropy, and internal energy.
\end{abstract}

\section{Introduction}

Transport in porous media takes place in a vast range of systems, natural as well as man-made. It is thus important to have a deep understanding of the relevant driving forces and their coupling, for instance to describe production of clean water \cite{rauter2021cassie, rauter2021thermo,qasim2019reverse}, CO$_2$ sequestration \cite{li2011pvtxy, ramdin2012state, boot2014carbon}, transport of oxygen, hydrogen, and water in fuel cell catalytic layers \cite{ sauermoser2020flow, wang2015theory}, and transport in lithium-ion battery electrodes and separators \cite{spitthoff2021peltier, gunnarshaug2020reversible,zhao2019review}.

\begin{figure}
    \centering
    \includegraphics[width=0.5\textwidth]{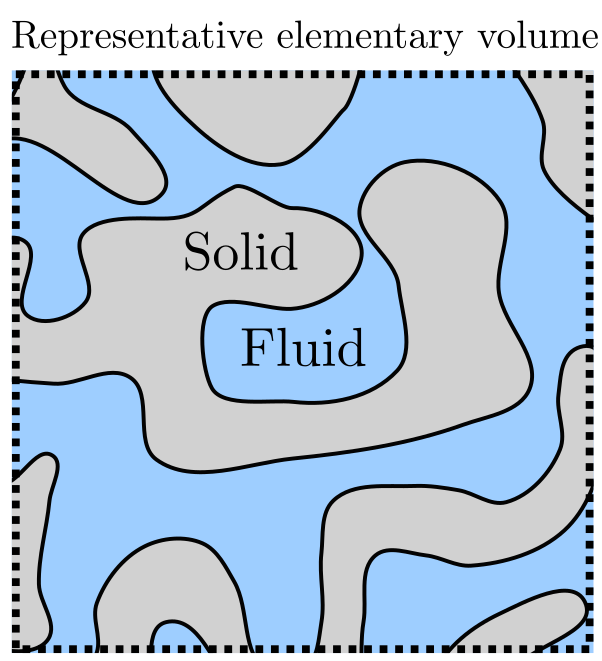}
    \caption{A single-phase and -component fluid in a porous medium. The fluid-solid interfaces can have complex shapes and the volumes can be on the nanoscale.}
    \label{fig:system}
\end{figure}

The long-range aim of this work is to obtain a general thermodynamic theory of transport of immiscible fluids in porous media on the macroscale \cite{hassanizadeh1990mechanics, Kjelstrup2018, Kjelstrup2019}. This theory must first describe the thermodynamic state of the fluids in the porous media. To do this we employ a bottom-up approach, a procedure that includes all details of the system on the nanoscale in the construction of a representative elementary volume (REV) \cite{bear1988dynamics, blunt2017multiphase, Bedeaux2020}. The procedure gives a coarse-grained description of the REV on the macroscale, or what is called the Darcy level \cite{bear1988dynamics, hassanizadeh1990mechanics,blunt2017multiphase,Kjelstrup2018,Kjelstrup2019}. A central issue is to find the pressure of the REV. Figure \ref{fig:system} illustrates the problem for the cases studied in this work. A fluid (blue) occupies the pores in a porous material (grey). The pores are so narrow that interactions between fluid and wall become significant, or in other words, that the fluid-wall surface energy becomes significant. But how can we define and determine the properties of the REV, for instance, the pressure? Can we find a representative elementary volume, for which this is possible? In this work, we aim to find answers to these questions.

In bulk fluids, the hydrostatic pressure is well defined, measurable, and well documented as the driving force of the fluid flow. In porous media, where fluids are confined by the pores, however, there is no consensus of neither the thermodynamic nor the mechanical definitions of the pressures. The microscopic mechanical pressure tensor is inherently ambiguous \cite{Irving1950,Schofield1982}. In this context, we make an emphasis on the milestone work of Israelachvili \cite{israelachvili2015intermolecular} who documented short- and long-range forces on fluid particles exerted by the surroundings, and on the thermodynamic analysis that was pioneered by Derjaguin \cite{Derjaguin1934}. 

We have previously proposed thermodynamic definitions for the pressure of heterogeneous media, applying Hill's idea of thermodynamics for small systems \cite{Kjelstrup2018, Kjelstrup2019, Galteland2019, Galteland2021, galteland2021legendre}. Hill's definitions have so far only been tested under simple conditions \cite{Rauter2020, Galteland2019, Galteland2021, Erdos2020}. They have not been studied for a variety of shapes and pore sizes. 

In Hill's thermodynamics of small systems or nanothermodynamics (See Bedeaux and Kjelstrup \cite{Bedeaux2020}), the REV of the porous medium is considered as an open system, controlled by the temperature, chemical potential, or pressure in the environment. The thermodynamic analysis is thus applied to a grand canonical ensemble of systems and a new variable is introduced, the replica energy, i.e. the energy needed to create one more small system in the ensemble. An adjusted Gibbs-Duhem equation (the Hill-Gibbs-Duhem's equation) appears. This opens up a new route to determine the REV pressure. 

Using this as a starting point in section \ref{sec:theory}, we shall find expressions for the so-called integral and differential pressures of the REV, from the system's replica energy. We present a new route to the pressure via the chemical potential of the reservoir in equilibrium with the REV. Molecular dynamics simulations, described in section \ref{sec:methods}, will be used to illustrate and verify the theoretical steps.

\section{Theory}
\label{sec:theory}

The systematic procedure of Hill consists of the three steps \cite{Bedeaux2020}, which we repeat to give an overview of the procedure. Concepts will be defined in the text that follows. 

\begin{enumerate}
    \item To start, define the items which make the system small in Hill's sense, by defining the relevant REV. Find the corresponding Hill-Gibbs equation. Next, define the environmental variables that control the system.
    \item From an analysis of the ensemble of small systems, find the system's replica energy in terms of its subdivision potential. 
    \item Derive the corresponding Hill-Gibbs-Duhem equation. This equation can be used to find the REV thermodynamic variables and their interrelations.
\end{enumerate}

\subsection{The general Hill-Gibbs equation of an ensemble of open porous media REVs}

We shall consider the REV of a porous medium, filled with a single-phase, single-component fluid, see figure \ref{fig:system}.  The REV volume is $V=V^f+V^s$, where $V^f$ is the volume of the fluid and $V^s$ is the volume of the solid. The porosity is the fraction of the fluid volume to the total volume, $\phi=V^f/V$. These are time-independent properties in the system. The REV is a small system in Hill's sense \cite{Hill1963, Bedeaux2020} because the fluid is confined; it does not have bulk-scale properties. The system is open for the supply of fluid particles and energy. The number of solid particles is fixed, however. 

The ensemble of $\mathcal{N}$ REVs has a total entropy $S_t$, total fluid volume $V^f_t$, total number of fluid particles $N^f_{t}$, total number of solid particles $N^s_{t}$ (with total volume $V^s_t$), and total fluid-solid surface area $A_t$. The total differential of the total internal energy of the ensemble of REVs is then
\begin{equation}
    \d U_t = T\d S_t-p^f\d V^f_t-p^s\d V^s_t + \mu^f\d N^f_{t} +\mu^s\d N^s_{t} +\gamma\d A_t +\varepsilon \d \mathcal{N}.
    \label{eq:total_internal_energy}
\end{equation}
This equation has been called the Hill-Gibbs equation \cite{Bedeaux2020, Galteland2021}. The temperature, fluid pressure, and solid pressures, chemical potentials, and surface tension are defined from the variables involved as
\begin{equation}
    \begin{split}
        T&\equiv\left(\frac{\partial U_t}{\partial S_t}\right)_{V^f_t,V^s_t,N^f_t,N^s_t,A_t,\mathcal{N}},\qquad
        p^f\equiv\left(\frac{\partial U_t}{\partial V^f_t}\right)_{S_t,V^s_t,N^f_t,N^s_t,A_t,\mathcal{N}},\\
        p^s&\equiv\left(\frac{\partial U_t}{\partial V^s_t}\right)_{S_t,V^f_t,N^f_t,N^s_t,A_t,\mathcal{N}},\qquad
        \mu^f\equiv\left(\frac{\partial U_t}{\partial N^f_t}\right)_{S_t,V^f_t,V^s_t,N^s_t,A_t,\mathcal{N}},\\
        \mu^s&\equiv\left(\frac{\partial U_t}{\partial N^s_t}\right)_{S_t,V^f_t,V^s_t,N^f_t,A_t,\mathcal{N}},\qquad
        \gamma\equiv\left(\frac{\partial U_t}{\partial A_t}\right)_{S_t,V^f_t,V^s_t,N^f_t,N^s_t,\mathcal{N}}.
    \end{split}
\end{equation}
The subdivision potential $\varepsilon$ is the property that deals with system smallness in particular. It is introduced with its conjugate variable, the number of REVs, $\mathcal{N}$. The subdivision potential is the internal energy required to add one more REV to the ensemble under the specified conditions,
\begin{equation}
    \varepsilon \equiv \left(\frac{\partial U_t}{\partial \mathcal{N}}\right)_{S_t, V^f_t, V^s_t,N^f_t, N^s_t, A_t}.
\end{equation}
When the subdivision potential differs from zero the system is small. This property will, as we shall see below, adjust the common variables like the pressure, and turn them into effective new variables. In the case of the pressure, the adjustment leads to the integral pressure, central for porous media. 

\subsection{The replica energy}

To find the properties of one system, we need to describe the thermodynamic state of a single-phase, single-component fluid in the REV. The fluid is free to move (the system is open), while the solid is not. The system exchanges fluid particles with the surroundings, but not solid particles. The variables that are controlled by contact with the environment are the temperature and the fluid chemical potential. Apart from these, as stated above,  the fluid volume, solid volume, surface area, and the number of solid particles are control variables too. The type of ensemble constituted by this set of variables is particularly suited for the transport of fluids through a porous medium.  

In order to change the set of variables in the general expression equation \ref{eq:total_internal_energy}, into the above preferred set of control variables, we express the additive variables in the original set of total variables by their controlled value times the number of replicas. This provides also the single system properties that we are after. The controlled fluid and solid volumes, and surface area per REV are $V^f_t = V^f \mathcal{N}$, $V^s_t = V^s\mathcal{N}$, and $A_t = A \mathcal{N}$, respectively. In addition we introduce the controlled number of solid particles per REV, $N_t^s=N^s\mathcal{N}$. By introducing these control variables into the Hill-Gibbs equation for the ensemble, we obtain 
\begin{equation}
    \d U_t = T\d S_t -p^f\mathcal{N}\d V^f-p^s\mathcal{N}\d V^s + \mu^f\d N_t^f + \mu^s\mathcal{N}\d N^s +\gamma\mathcal{N}\d A+X\d\mathcal{N},
    \label{eq:hill_gibbs2}
\end{equation}
where the last term is the replica energy of one small system
\begin{equation}
    X = \varepsilon-p^fV^f-p^sV^s+\gamma A+\mu^sN^s.
    \label{eq:Replica}
\end{equation}
The replica energy \cite{Hill1963, Hill1964} expresses in a simpler way, the energy required to add one more small system to the ensemble of systems under the conditions controlled. The combination of terms can define the integral pressure minus the integral solid chemical potential times number of solid particles \cite{Galteland2021},
\begin{equation}
     -\hat{p}V +\hat{\mu}^sN^s \equiv X,
\end{equation}
however, we need an additional equation to determine both $\hat{p}$ and $\hat{\mu}^s$ separately. Through the introduction of an ensemble of the systems, we have achieved that the internal energy, here $U_t$, is an Euler homogeneous of the first order in the number of REVs. This gave Hill the motivation for the use of an ensemble, and to define the conjugate variables $\varepsilon$ and $\mathcal{N}$. 

\subsection{The Hill-Gibbs equation for a single small system}

We are now in a position to integrate the total differential of the total internal energy, see equation \ref{eq:hill_gibbs2}, at constant $T, V^f, V^s, \mu^f, N^s$, and $A$. This gives,
\begin{equation}
    U_t = TS_t+\mu^fN^f_t+X\mathcal{N}.
\end{equation}
By introducing the REV average properties, we obtain the internal energy of one REV,
\begin{equation}
    U = TS+\mu^fN^f+X.
    \label{eq:internal_energy}
\end{equation}
The total differential of the internal energy of one REV is
\begin{equation}
    \d U = T\d S -p^f\d V^f-p^s\d V^s+\mu^f\d N^f+\mu^s\d N^s + \gamma \d A.
\end{equation}
By differentiating the internal energy of the REV and using its total differential, we obtain the total differential of the replica energy
\begin{equation}
    \d X = \d(-\hat{p}V+\hat{\mu}^sN^s) = -S\d T -p^f\d V^f-p^s\d V^s-N^f\d \mu^f+\mu^s\d N^s +\gamma \d A.
    \label{eq:HGD}
\end{equation}
The equation is the outcome of step 2 in Hill's procedure presented in the introduction to this section. The equation can be seen as an extension of Gibbs-Duhem's equation, so we have called it Hill-Gibbs-Duhem's equation \cite{Bedeaux2020}. Applications can now be specified. 

The partial derivatives of the replica energy follow 
\begin{equation}
    \begin{split}
        S&=-\left(\frac{\partial X}{\partial T}\right)_{V^f,V^s,\mu^f,N^s, A},\qquad
        p^f=-\left(\frac{\partial X}{\partial V^f}\right)_{T,V^s,\mu^f,N^s, A},\\
        p^s&=-\left(\frac{\partial X}{\partial V^s}\right)_{T,V^f,\mu^f,N^s, A},\qquad
        N^f=-\left(\frac{\partial X}{\partial \mu^f}\right)_{T,V^f,V^s,N^s, A},\\
        \mu^s&=\left(\frac{\partial X}{\partial N^s}\right)_{T,V^f,V^s,N^f,A},\qquad
        \gamma=\left(\frac{\partial X}{\partial A}\right)_{T,V^f,V^s,N^f,\mu^s}.
    \end{split}
\end{equation}
With this set of equations we can calculate all the necessary REV properties of a porous medium. We shall concentrate on the integral fluid pressure and the route to this quantity via the chemical potential.  

\subsection{The integral pressure and the chemical potential of the solid in the REV}

We apply here the conditions of constant temperature, fluid volume, solid volume, number of solid particles, and surface area. The Hill-Gibbs-Duhem's equation reduces to
\begin{equation}
    \d X = \d(-\hat{p}V+\hat{\mu}^sN^s) =   -N^f\d \mu^f  
    \label{eq:HGD2}
\end{equation}
or, after dividing by $V$, the volume of the REV, 
\begin{equation}
    \d x = \d(-\hat{p}+ \hat{\mu}^s\rho^s) =   -\rho^f\d\mu^f  
    \label{eq:HGD3}
\end{equation}
Where the density of the fluid in the REV is $\rho^f = N^f/V$ and the density of the solid is $\rho^s = N^s/V$. These densities are of the total REV volume $V$, and not the fluid volume $V^f$ or solid volume $V^s$. The difference of the replica energy density can be calculated from
\begin{equation}
    x -x_\infty= -\int_{-\infty}^{\mu^f}\rho^f\d\mu'^f,
    \label{eq:replica_energy_density}
\end{equation}
The replica energy density is zero as the fluid chemical potential approaches minus infinity. The replica energy density depends on two unknown variables, $\hat{p}$ and $\hat{\mu}^s$. To proceed we need to know more about these variables. 

\subsubsection{Constant integral pressure across boundary}

The integral pressure of the REV can be obtained from equation \ref{eq:HGD3}
\begin{equation}
    \d\hat{p} = \rho^f\d\mu^f + \rho^s\d\hat{\mu}^s.
    \label{eq:dp_hat}
\end{equation}
The fluid in the porous medium is in equilibrium with its environment, here the bulk phase fluid that surrounds the system (denoted $b$). The environment has the same temperature. The integral pressure was observed to be constant across the phase boundary inside a pore \cite{Rauter2020}. We shall therefore make the assumption that also in this case; 
\begin{equation}
     \hat{p} = p^b 
    \label{eq:loce}
\end{equation}
where $p^b$ is the bulk pressure of a fluid in equilibrium with the porous medium. For an ideal gas in a cubic confinement it has been shown that the integral pressure is equal to the bulk pressure \cite{braaten2021small}. If this applies for this system, we also have
\begin{equation}
    \hat{p} = \int_{-\infty}^{\mu^f}\rho^b\d\mu'^f,
\end{equation}
where $\rho^b$ is the fluid number density in the bulk phase. The integral pressure is zero as the fluid chemical potential approaches minus infinity. As a consequence, the integral pressure depends only on the fluid chemical potential and temperature. The integral chemical potential of the solid in the REV is then
\begin{equation}
    \hat{\mu}^s = \frac{1}{\rho^s}\int_{-\infty}^{\mu^f}(\rho^b-\rho^f)\d \mu'^f.
    \label{eq:solid_chemical_potential}
\end{equation}
The solid chemical potential is zero at minus infinite fluid chemical potential. The entropy density of the REV is
\begin{equation}
    s = \frac{S}{V} = \frac{1}{T}( u - \mu^f\rho^f-x)
    \label{eq:entropy}
\end{equation}
where we used equation \ref{eq:internal_energy} and introduced the replica energy density, $x$.

\subsection{Pressures and surface tension}
There are three pressures; the integral, fluid and solid pressures. The latter two are differential pressures. The integral pressure is a combination of the differential pressures and the (differential) surface tension,
\begin{equation}
    \hat{p} = p^f\phi+p^s(1-\phi)-\gamma A/V = p^b.
    \label{eq:phat}
\end{equation}
We assume the integral pressure to be equal everywhere in equilibrium and also equal to the bulk pressure of a bulk fluid in equilibrium with the porous medium. If we also assume the distances between solid surfaces to be large and their curvature to be small, we can approximate the fluid pressure to be equal to the bulk pressure \cite{Galteland2021}. In that case, the solid pressure is equal to
\begin{equation}
    p^s = p^b + \frac{\gamma A}{V^s}.
    \label{eq:solid_pressure}
\end{equation}

\section{Simulation details}
\label{sec:methods}
\begin{figure}
    \centering
    \includegraphics[width=\textwidth]{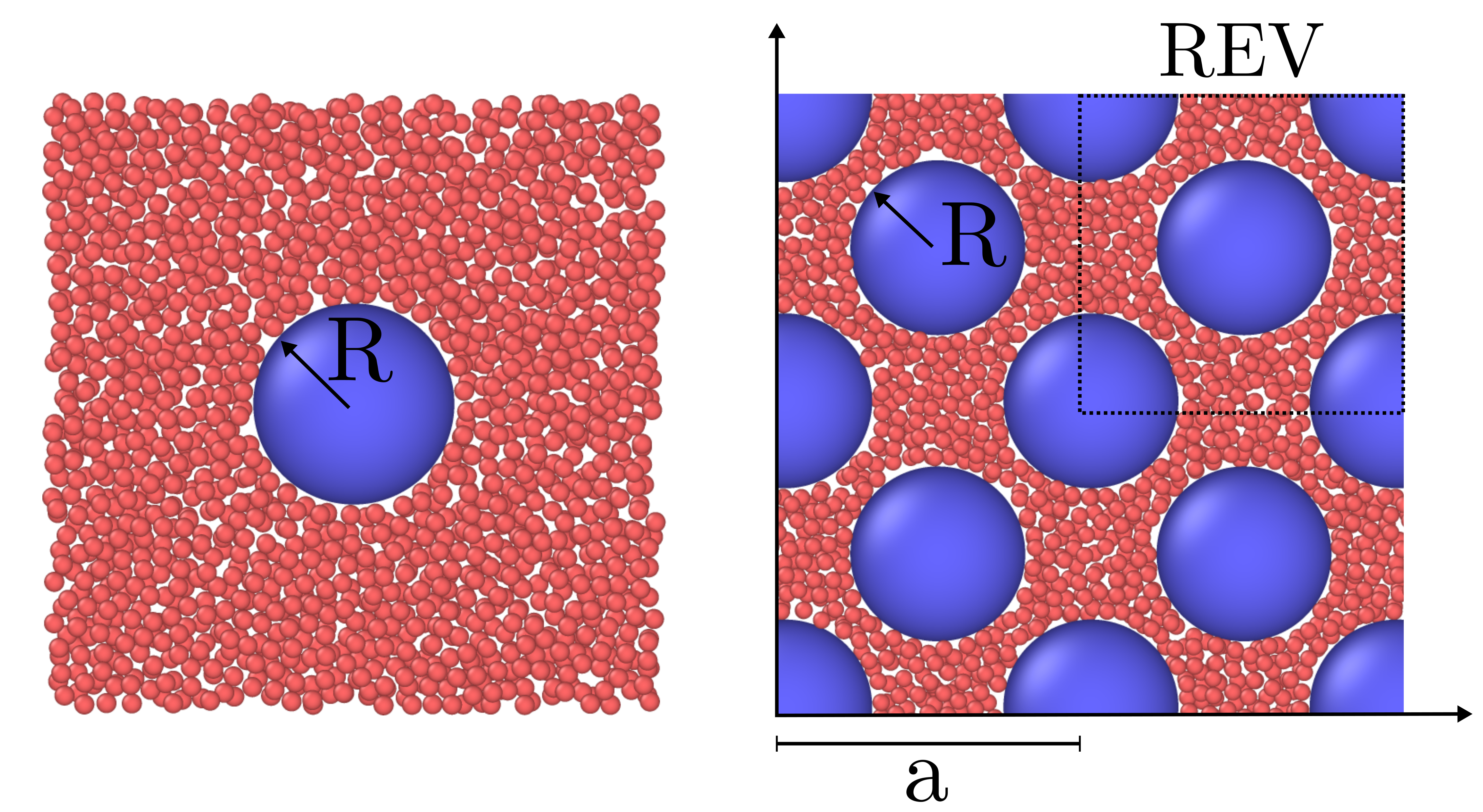}
    \caption{Visualization of a two of the simulated systems. Fluid particles (red) and solid particles (blue) are visualised with OVITO \cite{Stukowski2009}. Only the particles in a slab centered on the solid particles of thickness $2.5\sigma$ is shown. The fluid chemical potential is $\mu^f=\zeta$ and the radius of the solid particles is $R=5\sigma$. Left: Single solid particle. Right: An fcc lattice with lattice constant $a=17.5\sigma$ and porosity $\phi=0.61$.}
    \label{fig:simulation}
\end{figure}

Systems of fluid and solid particles were investigated with molecular dynamics simulations using LAMMPS \cite{Plimpton1995,thompson2021lammps}. The three different systems have been simulated: A bulk fluid, a single solid particle surrounded by fluid particles, and a face-centered cubic (fcc) lattice of solid particles filled with fluid particles in the pore space. The two latter systems are illustrated in figure \ref{fig:simulation}. The systems were simulated in the grand canonical ensemble and had periodic boundary conditions in all directions. The bulk fluid was simulated to calculate the bulk pressure as a function of the fluid chemical potential. The single solid particle surrounded by fluid particles will be considered as if the lattice constant of the fcc lattice is large. This system was simulated to calculate the fluid-solid surface tension when the solid particles are far apart. The thermodynamic properties of the fluid in the fcc lattice were calculated as a function of the fluid chemical potential. 

The temperature was controlled by using a Nosé-Hoover thermostat \cite{nose1984unified, hoover1985canonical} adapted by Shinoda \textit{et. al.} \cite{shinoda2004rapid}. The fluid chemical potential was controlled by using grand canonical Monte Carlo insertions and deletions of fluid particles \cite{Frenkel2001}. A particle is inserted at random position in the simulation box with a probability
\begin{equation}
    \text{acc}(N^f\rightarrow N^f+1) = \min\left\{1,\quad \frac{V}{\Lambda^3(N^f+1)}\exp[\beta(\mu-E_p(N^f+1)+E_p(N^f)] \right\}.
\end{equation}
A random particle is removed from the simulation box with a probability
\begin{equation}
    \text{acc}(N^f\rightarrow N^f-1) = \min\left\{1,\quad \frac{\Lambda^3(N^f+1)}{V}\exp[-\beta(\mu+E_p(N^f-1)-E_p(N^f)] \right\}.
\end{equation}
Where $\Lambda=\sqrt{h^2/(2\pi m k_\text{B}T)}$ is the de Broglie thermal wavelength, $h$ is the Planck constant, $E_p$ is the potential energy and $\beta=1/(k_\text{B}T)$ where $k_\text{B}$ is Boltzmanns constant.

\subsection{Simulation procedure}

In the bulk system, the fluid chemical potential was varied in the range $\mu^f\in[-7,1]$, in which the fluid density varied from a dilute gas to a dense liquid. The system was initialized with an empty cubic simulation box of volume $V=(40\sigma)^3$, and the grand canonical Monte Carlo technique was used to insert particles. The simulation was run until the average number of fluid particles was no longer increasing. The bulk fluid number density $\rho^b=N^f/V$ and pressure $p=(P_{xx}+P_{yy}+P_{zz})/3$ were calculated as a function of the fluid chemical potential for $10^7$ steps. The calculation of the mechanical pressure tensor $P_{\alpha\beta}$ is described below in subsection \ref{sec:mechanical_pressure_tensor}.

The system of the single solid particle was simulated by first placing the solid particle in the center of the cubic simulation box of volume $V=(30\sigma)^3$. The radius of the solid particle was $R=5\sigma$, where $\sigma$ is the diameter of the fluid particle. The fluid chemical potential was varied in the range $\mu\in[-7, 1]\zeta$, where $\zeta$ is the minimum of the pair-wise interaction. The system was initialized by placing the fluid particles in an fcc lattice with a fluid number density approximately equal to the bulk fluid number density around the solid particle at a given fluid chemical potential. This was done to reduce the computational time. The simulation was run until the average number of fluid particles no longer increased, after which it was run for an additional $10^7$ steps to calculate the surface tension as a function of the fluid chemical potential. The calculation of the surface is described below in section \ref{sec:mechanical_pressure_tensor}.

The fcc lattice of solid particles was fixed in space with lattice constant in the range $a\in[15,30]\sigma$, the radius of the solid particles was $R=5\sigma$. The simulation box lengths were $L=2a$, total volume $V=8a^3$, number of solid particles $N^s = 32$, and solid volume $V^s = 128\pi R^3/3$. We have previously found that the smallest REV of such a system is a quarter unit cell \cite{Galteland2019}. The fluid particles were inserted in the pore space in between the solid particles. The porosity of the system was
\begin{equation}
    \phi = \frac{V^f}{V}=1-\frac{16\pi R^3}{3a^3},
\end{equation}
where the radius of the solid particles was $R=5\sigma$ given that the lattice constant is $a\geq2\sqrt{2}R\approx14\sigma$. For lower values of $a$ the solid particles overlap, however, we do not look at such systems. The porosity varied from approximately $0.38$ to $0.92$. The fluid chemical potential, number of solid particles, temperature, and volume were controlled. 

The system was initialized by placing the solid particles in an fcc lattice with lattice constant $a$ and fluid particles also in an fcc lattice with a number density approximately equal to $\rho^b(\mu^f)$ around the solid particles. The simulation was run until the number of fluid particles was constant, after which the simulation was run for an additional $10^7$ timesteps to calculate thermodynamic properties.

\subsection{The Lennard-Jones/spline potential}
The particles interacted with the Lennard-Jones/spline (LJ/s) potential \cite{Hafskjold2019}, which is equal to the Lennard-Jones potential up to the inflection point $r_s$ shifted by a hard-core diameter $d$. At $r_s$ a third degree polynomial is fitted for the potential to be zero at the cut-off $r_c$ and the force and potential to be continuous at the inflection point $r_s$ and cut-off $r_c$. See the works by Hafskjold \textit{et al.} \cite{Hafskjold2019} and Kristiansen \cite{kristiansen2020transport} for details on the properties of the LJ/s potential. The LJ/s potential is,
\begin{equation}
    \nu^\text{LJ/s}(r)= 
    \begin{cases} 
    4\zeta\left[\left(\frac{\sigma}{r-d}\right)^{12}-\left(\frac{\sigma}{r-d}\right)^{6}\right] &\text{if } r<r_s,\\
    a(r-r_c)^{2}+b(r-r_c)^{3} &\text{if } r_s<r<r_c,\\
    0 &\text{else}
    \end{cases}
\end{equation}
where $\zeta$ is the minimum of the interaction potential. This symbol is used to avoid confusion with the subdivision potential $\varepsilon$. The soft-core diameter is $\sigma$, $d$ is the hard-core diameter, $r=|\bm{r}_j-\bm{r}_i|$ is the distance between particle $i$ and $j$. The distance $r=\sigma+d$ is the smallest distance where the potential is zero. The Lennard-Jones/spline interaction potential was equal for all particle pairs, but was shifted with hard-core diameter $d$. There were three particle pair interactions, fluid-fluid, fluid-solid, and solid-solid, and only the hard-core diameter varied between them. The solid-solid interaction we set to zero. The hard-core diameter was $d_\text{ff}=0$, $d_\text{fs}=4.5\sigma$, and $d_\text{ss}=9\sigma$ for the fluid-fluid, fluid-solid, and solid-solid interactions, respectively. The solid particle radius was defined as $R\equiv (d_\text{ss}+\sigma)/2 = 5\sigma$. Other definitions of the radii are possible, for example based on Gibbs dividing surface or the surface of tension. The parameters $a$, $b$, $r_s$ and $r_c$ were determined such that the potential and the force were continuous at $r=r_s$ and $r=r_c$. The masses of fluid particles were $m$, and the solid particles were considered to have infinite mass as they were fixed in space.

\subsection{Internal energy density}

The internal energy density was calculated as the sum of the kinetic and potential energy densities,
\begin{equation}
    u = \frac{1}{V}\left(\sum_{i=1}^N\sum_{j>i}^N u^\text{LJ/s}(r_{ij}) + \frac{1}{2}\sum_{i=1}^Nm_i(\bm{v}_i\cdot\bm{v}_i)\right)
\end{equation}
where $V$ is the total volume, $N$ is the total number of particles, and $\bm{v}_i$ is the velocity of particle $i$. The solid particles did not contribute to the kinetic energy, as their velocities were controlled to be zero. The internal energy density was used to calculate the entropy density, see equation \ref{eq:entropy}.

\subsection{The mechanical pressure tensor}
\label{sec:mechanical_pressure_tensor}

The mechanical pressure tensor was calculated in spherical and Cartesian coordinates, see Ikeshoji \textit{et.al.} for details \cite{Ikeshoji2003}. It was calculated in spherical coordinates for the single solid particle surrounded by fluid, and in Cartesian coordinates for the bulk fluid and the fcc lattice of solid particles with inserted fluid particles. For the bulk fluid and fcc lattice, the pressure was calculated for the whole simulation box, with sides $L\in[30, 60]\sigma$ and interaction cut-off $r_c\approx 1.74\sigma$. While for the single solid particle it was calculated in spherical shell subvolumes with the origin at the center of the solid particle. This was done to calculate the surface tension.

To avoid confusion between the thermodynamic and mechanical pressures, we will use lower case $p$ for all thermodynamic pressures and upper case $P$ for mechanical pressures. We make this distinction because there is no consensus on how, if possible, to connect the two for heterogeneous media \cite{Long2011, Dijk2020, Long2020, Galteland2021}.

The mechanical pressure tensor can be written as the sum of the ideal gas contribution and a virial contribution,
\begin{equation}
    P_{\alpha\beta} = \rho k_\text{B}T\delta_{\alpha\beta}+P_{\alpha\beta}^v,
\end{equation}
where $\delta_{\alpha\beta}$ is the Kronecker delta and the subscripts give the components of the tensor. The virial contribution is due to particle pair interaction, and is calculated as a sum over all particle pairs. For a subvolume $V_k$ the virial contribution is
\begin{equation}
    P_{\alpha\beta}^v = - \frac{1}{V_k}\sum_{i=1}^N\sum_{j>i}^Nf_{ij,\alpha}\int_{C_{ij}\in V_k}\d l_\beta.
\end{equation}
Where $N$ is the total number of particles and $f_{ij,\alpha}$ is the $\alpha$-component of force acting on particle $i$ due to particle $j$. The line integral is along the part of the curve $C_{ij}$ that is contained in the subvolume $V_k$. The virial contribution is inherently ambiguous as any continuous curve that starts at the center particle $i$ and ends at the center particle $j$ is permitted \cite{Irving1950, Harasima1958, Schofield1982}. In this work, we have used the Irving-Kirkwood curve, which is the straight line from the center of particle $i$ to the center of particle $j$.

The Cartesian mechanical pressure tensor was calculated for the whole simulation box. The line integral in the virial contribution reduces it to
\begin{equation}
    \label{eq:mechanical_pressure}
    P_{\alpha\beta}^v = - \frac{1}{V_k}\sum_{i=1}^N\sum_{j>i}^Nf_{ij,\alpha}r_{ij,\beta}.
\end{equation}
where $r_{ij}$ is the line for the center of particle $i$ to the center of particle $j$.

The pressure is calculated as the mean of the diagonal components of the Cartesian mechanical pressure tensor. The bulk pressure is
\begin{equation}
    p^b = \frac{1}{3}(P_{xx}+P_{zz}+P_{yy}).
\end{equation}
This was also calculated for the fluid in the fcc lattice of solid spheres. However, we have not equated it to a thermodynamic property.

The fluid-solid surface tension was calculated for the single-sphere simulation case from the spherical mechanical pressure tensor,
\begin{equation}
    \gamma = \frac{1}{R^2}\int_R^{r_0}(P_N-P_T) r^2\d r.
    \label{eq:surface_tension}
\end{equation}
Where $R=5\sigma$ is the solid particle radius, and $r_0=14\sigma$ is a position in the fluid far away from the fluid-solid surface. The normal component is $P_N = P_{rr}$ and the tangential component is $P_T = (P_{\phi\phi}+P_{\theta\theta})/2$. This is the fluid-solid surface tension of the porous medium given that the surfaces are sufficiently far apart. This will be used to calculate the solid pressure, see equation \ref{eq:solid_pressure}.

\section{Results and discussion} 

\subsection{Single solid particle surrounded by fluid}
\begin{figure}
    \centering
    \includegraphics[width=\textwidth]{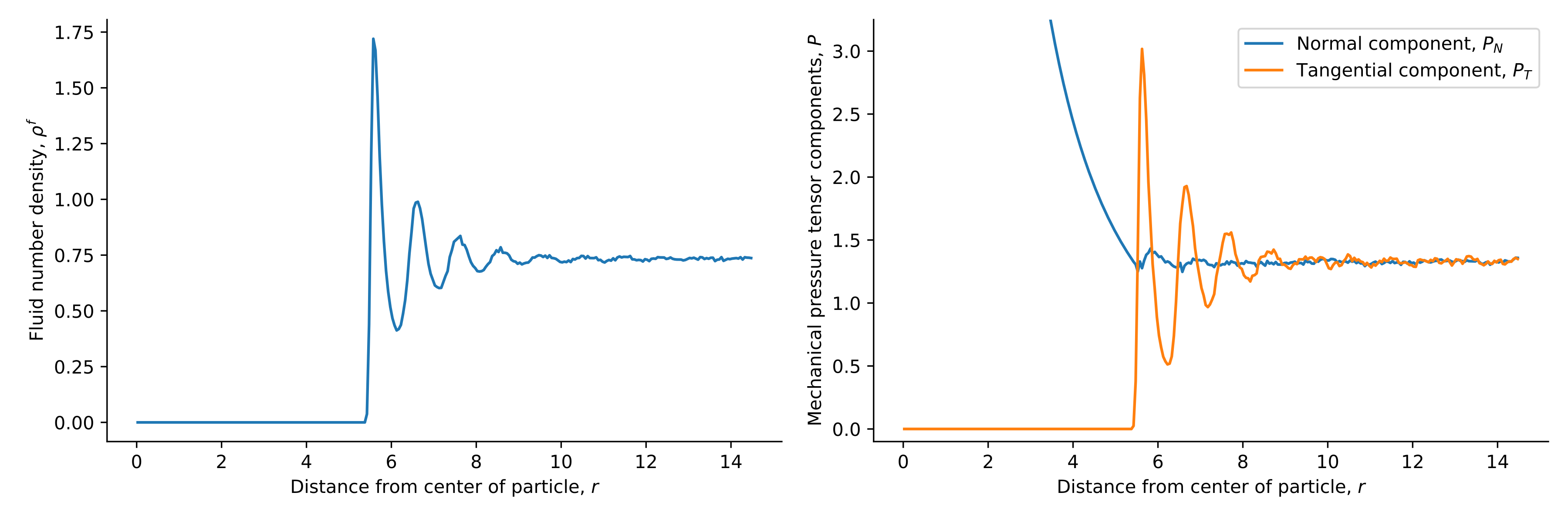}
    \caption{Local fluid number density (left) and normal and tangential pressure tensor components (right) in spherical shells as a function of the distance from the center of the spherical particle in the single-sphere simulation case. The fluid chemical potential is fixed to $\mu^f=1\zeta$ and the particle radius is $R=5\sigma$.}
    \label{fig:profile}
\end{figure}

For the single solid sphere, we show the 
local fluid number density and mechanical pressure tensor components of the surrounding fluids in figure \ref{fig:profile}. The fluid chemical potential is fixed at the minimum of the pair-wise interaction potential, $\zeta$, giving $\mu^f=1\zeta$. At this fluid chemical potential, the bulk density is $\rho^b = 0.8\sigma^{-3}$. The fluid particles pack in layers close to the surface of the solid particle of radius $R=5\sigma$, as reflected in the density variations in figure \ref{fig:profile}. The tangential component of the mechanical pressure tensor, shown in orange to the right in the figure,  has the same variation. The normal and tangential components were used to calculate the surface tension, shown in figure \ref{fig:surface_tension} (see below). In this calculation, we do not equate the solid pressure to the mechanical pressure in any way. Inside the solid particle, the tangential component is zero. This is because the cut-off of the fluid-fluid interaction is relatively short, meaning that no fluid-fluid interaction contributes to the pressure inside the solid sphere. It is only the fluid-solid interactions that contribute to the pressure inside, and these contribute only to the normal component. The normal component is proportional to $r^{-2}$ as a consequence, see Fig.\ref{fig:profile}.  This follows from mechanical equilibrium in spherical coordinates,
\begin{equation}
    P_T(r) = P_N(r) + \frac{r}{2}\frac{\partial P_N}{\partial r}.
\end{equation}
We used mechanical equilibrium and found that the mechanical pressure calculation in spherical coordinates is consistent with this. 

The surface tension was calculated for a single solid sphere surrounded by fluid particles, as a function of the fluid chemical potential. This is presented in figure \ref{fig:surface_tension}. It was calculated from the components of the mechanical pressure tensor, see equation \ref{eq:surface_tension}. Fluid particles have reduced contact with each other and with the solid in the lower chemical potential-regime ($\mu^f\in[-7,-3]\zeta$), which is characterized by low densities. The surface tension is accordingly very small. The surface tension between fluid and solid increases sharply at a chemical potential of around $\mu^f = -2.3\zeta$. The increase indicates increasing interactions between particles (fluid-fluid and fluid-solid). For a chemical potential below $\approx$ 2.3$\zeta$, the fluid is more vapor-like, while above this value, it is more liquid-like.

\begin{figure}
    \centering
    \includegraphics[width=0.75\textwidth]{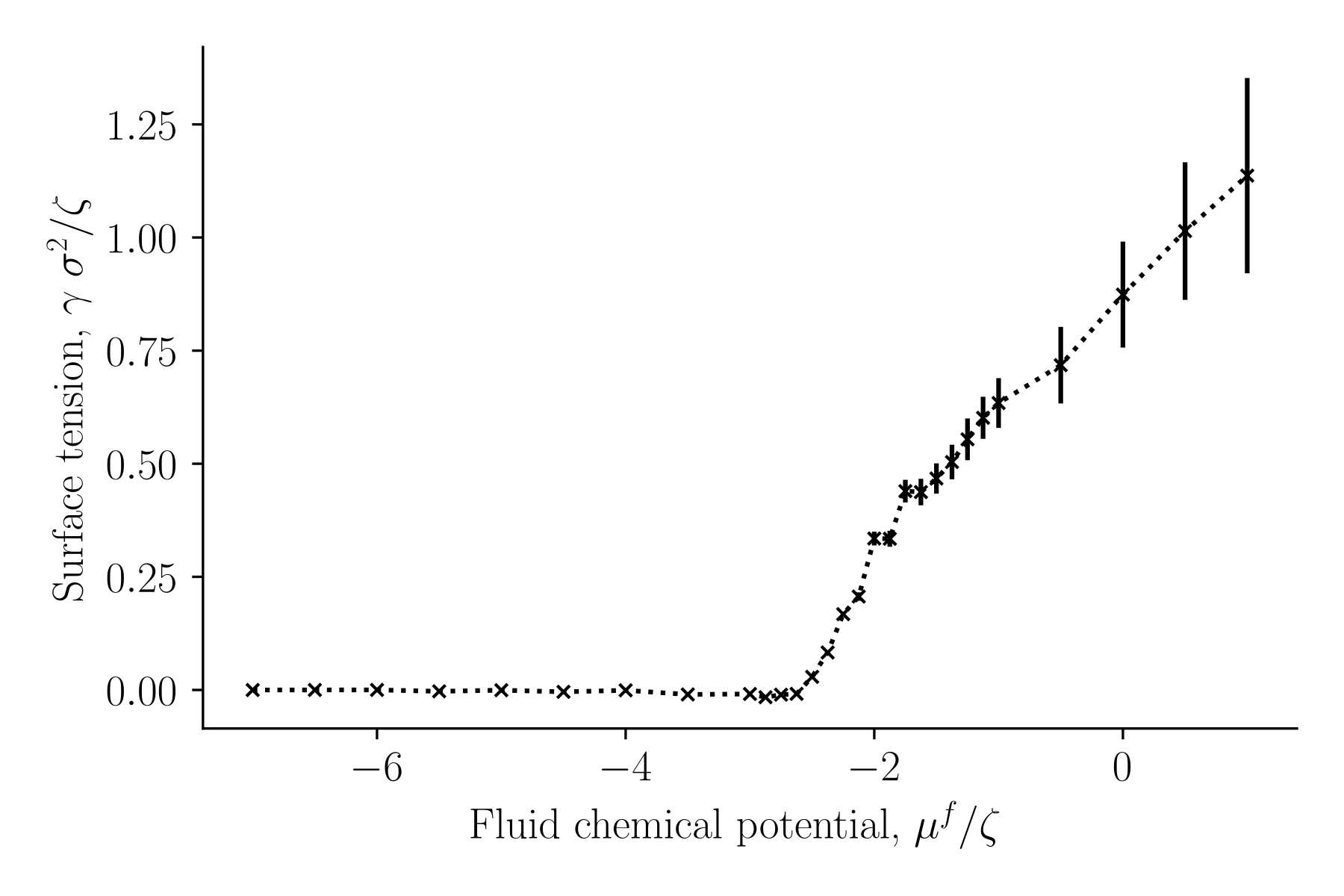}
    \caption{Surface tension for the single-sphere case $\gamma$ as a function of fluid chemical potential $\mu^f$.}
    \label{fig:surface_tension}
\end{figure}

\begin{figure}
    \centering
    \includegraphics[width=0.75\textwidth]{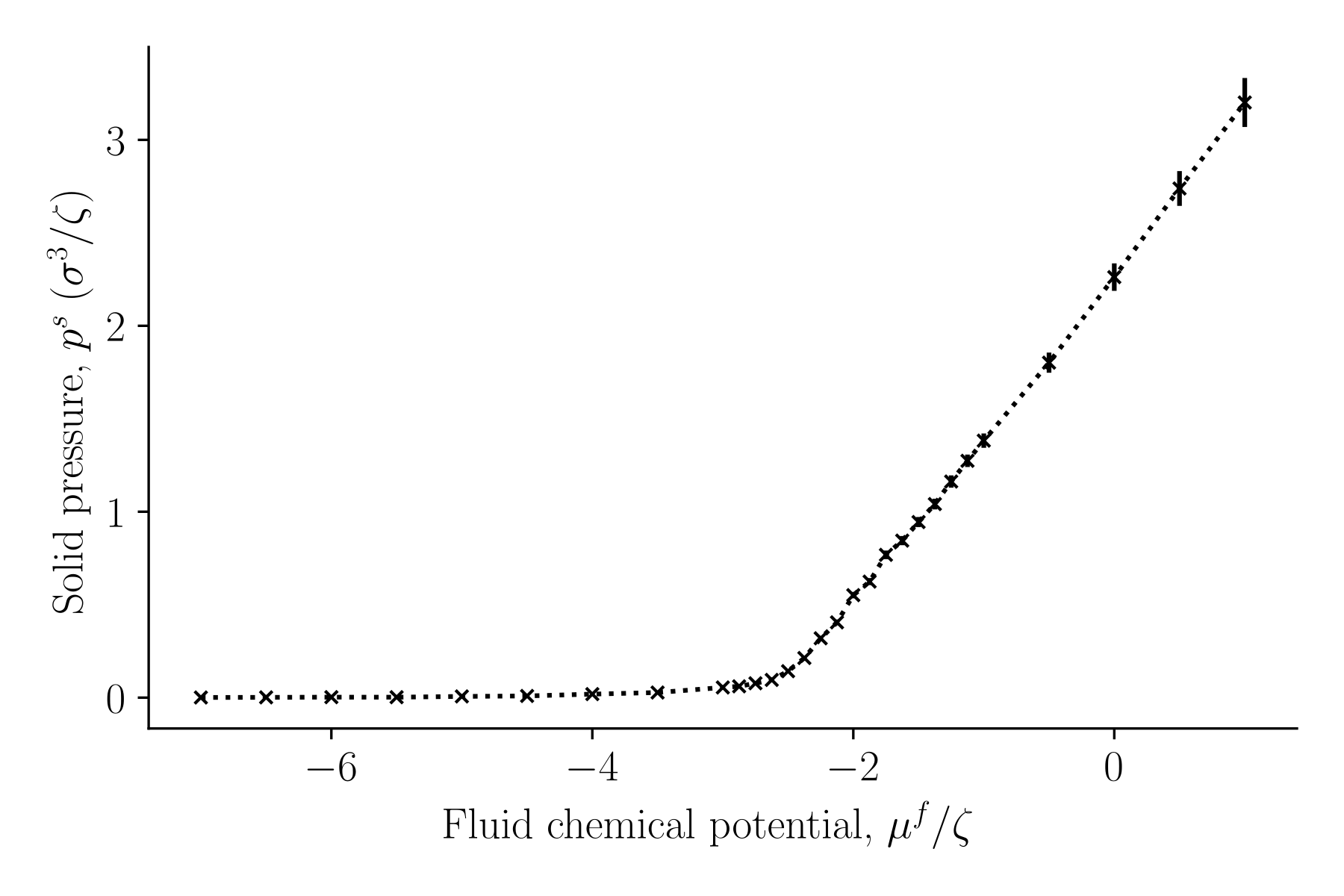}
    \caption{The solid pressure, $p^s$, in a single sphere, calculated from equation \ref{eq:solid_pressure} as a function of the fluid chemical potential $\mu^f$ of the surrounding fluid.}
    \label{fig:solid_pressure}
\end{figure}

In figure \ref{fig:solid_pressure} the corresponding solid pressure is shown as a function of fluid chemical potential for a single-sphere surrounded by fluid particles. It is calculated from the surface tension and bulk pressure, see equation \ref{eq:solid_pressure}. The bulk pressure monotonically increases with the fluid chemical potential.

So far, the variations are all expected from mechanical equilibrium and standard thermodynamics as given by the Young-Laplace equation. It is reasonable that the pressure of the solid increases monotonously, following the variation in the surface tension. 

\subsection{The fluid number density and the replica energy density}

\begin{figure}
    \centering
    \includegraphics[width=0.75\textwidth]{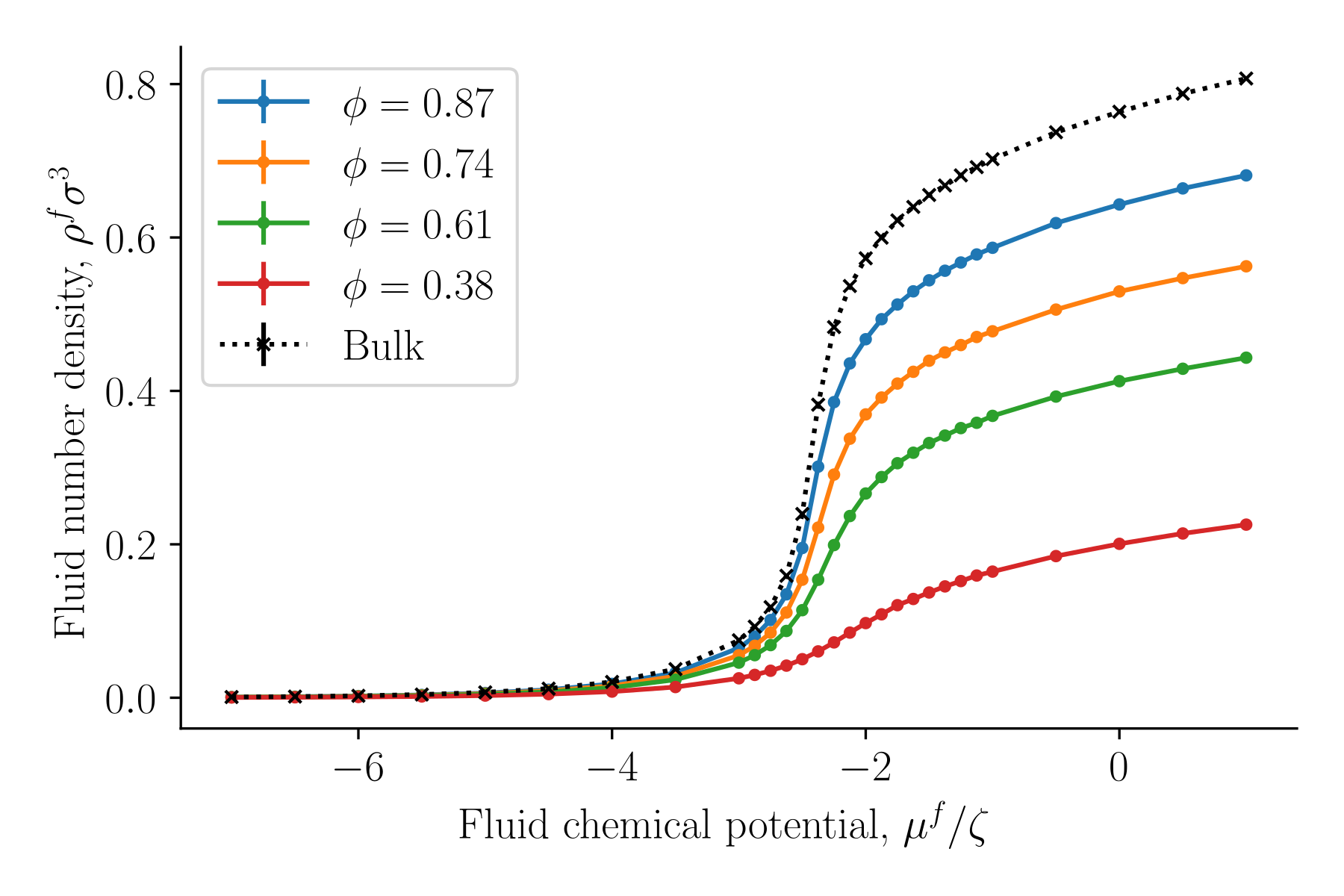}
    \caption{Fluid number density $\rho^f = N^f/V$ as a function of fluid chemical potential $\mu^f$ at varying porosity $\phi=V^f/V$.}
    \label{fig:fluid_density}
\end{figure}

Figure \ref{fig:fluid_density} presents the fluid number density $\rho^f=N^f/V$, where $V$ is the total REV volume, as a function of the controlled fluid chemical potential $\mu^f$ for various porosities $\phi$ of the fcc lattice. Results shown as black crosses represent bulk fluid or the limit where the porosity approaches unity. The fluid number density starts at approximately zero density for small fluid chemical potentials (a dilute gas-like phase) and converges to a density of a dense liquid-like phase. The fluid number density decreases with decreasing porosity, which is mainly because of decreased fluid volume compared to total volume. The fluid particles form layers on the solid surface, as was seen in figure \ref{fig:profile}.

\begin{figure}
    \centering
    \includegraphics[width=0.75\textwidth]{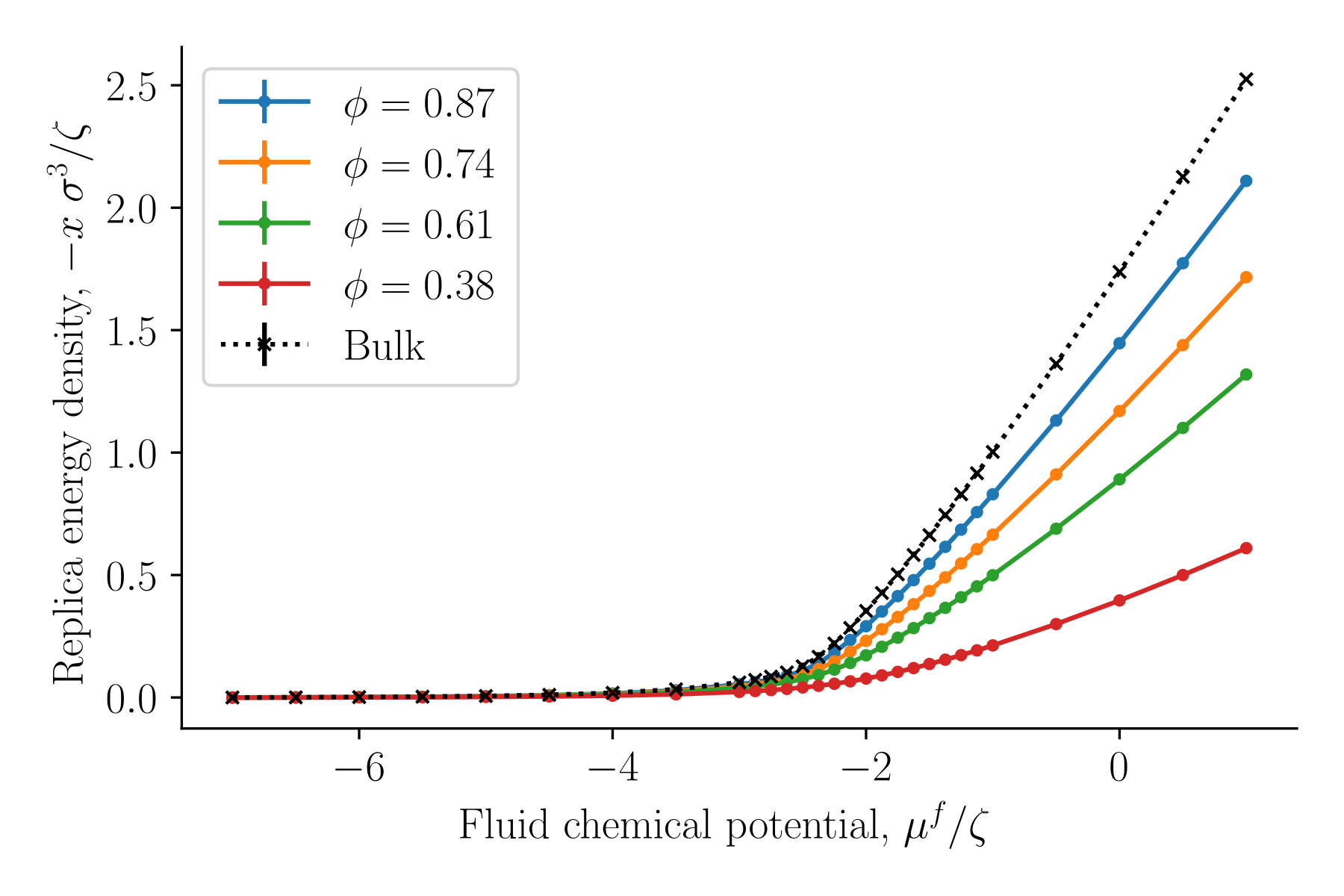}
    \caption{The negative replica energy density $-x=-X/V$ as a function of the fluid chemical potential $\mu^f$ at varying porosities $\phi=V^f/V$.}
    \label{fig:replica_energy_density}
\end{figure}

The replica energy density is a characteristic property of the small system. It is shown for the first time for a regular fcc-lattice in figure \ref{fig:replica_energy_density}. The property was obtained as the integral over the fluid density, of the fluid chemical potential, see equation \ref{eq:replica_energy_density}. The replica energy density approaches the bulk fluid value (black crosses) in the thermodynamic limit of increasing porosity, as expected. We see a similar development in the replica energy density, as seen in the curve for the pressure of the solid around the single sphere. The replica energy is negative and increases with decreasing porosity. When compared with other thermodynamic properties like the internal energy (see below), the value is sizable. The system is small in the sense that it has non-negligible replica energy. 
  
\begin{figure}
    \centering
    \includegraphics[width=0.75\textwidth]{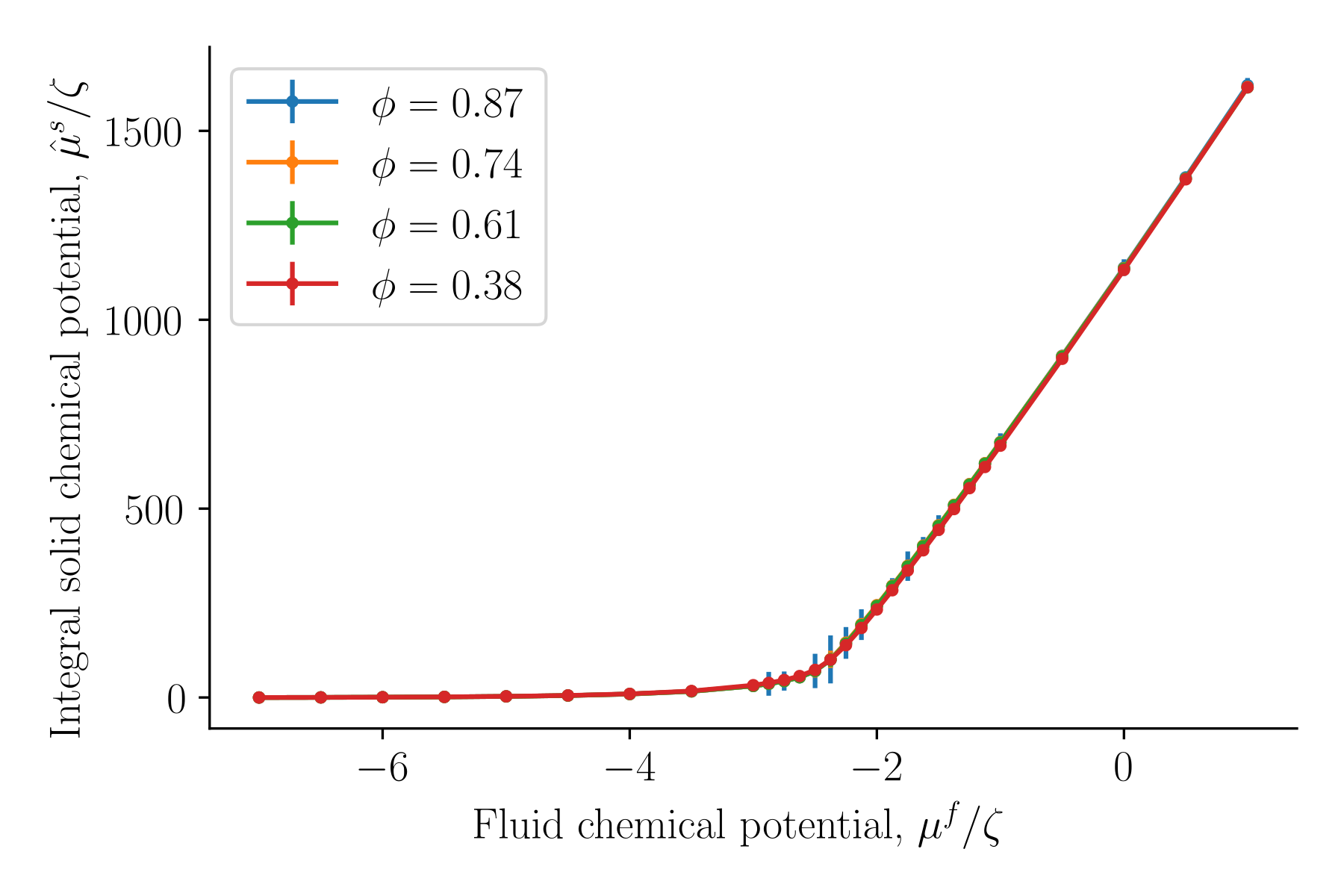}
    \caption{Integral solid chemical potential as a function of the fluid chemical potential $\mu^f$ at varying porosities $\phi=V^f/V$.}
    \label{fig:solid_chemical_potential}
\end{figure}

The integral solid chemical potential was next calculated from the fluid number density of the REV and the bulk, see equation \ref{eq:solid_chemical_potential}. It is presented in figure \ref{fig:solid_chemical_potential} as a function of the fluid chemical potential. The integral solid chemical potential is much larger than the fluid chemical potential because each solid particle has a radius ten times larger than a fluid particle. This implies that each solid particle interacts with more particles than each fluid particle. The energy required to add one more solid particle is very large compared to adding one more fluid particle. Interestingly, the integral solid chemical potential is independent of the porosity, it is a function of fluid number density and fluid chemical potential. We interpret this to mean that the solid particle are sufficiently far away from each other, such that the integral solid chemical potential is unaffected by the porosity. 

\begin{figure}
    \centering
    \includegraphics[width=0.75\textwidth]{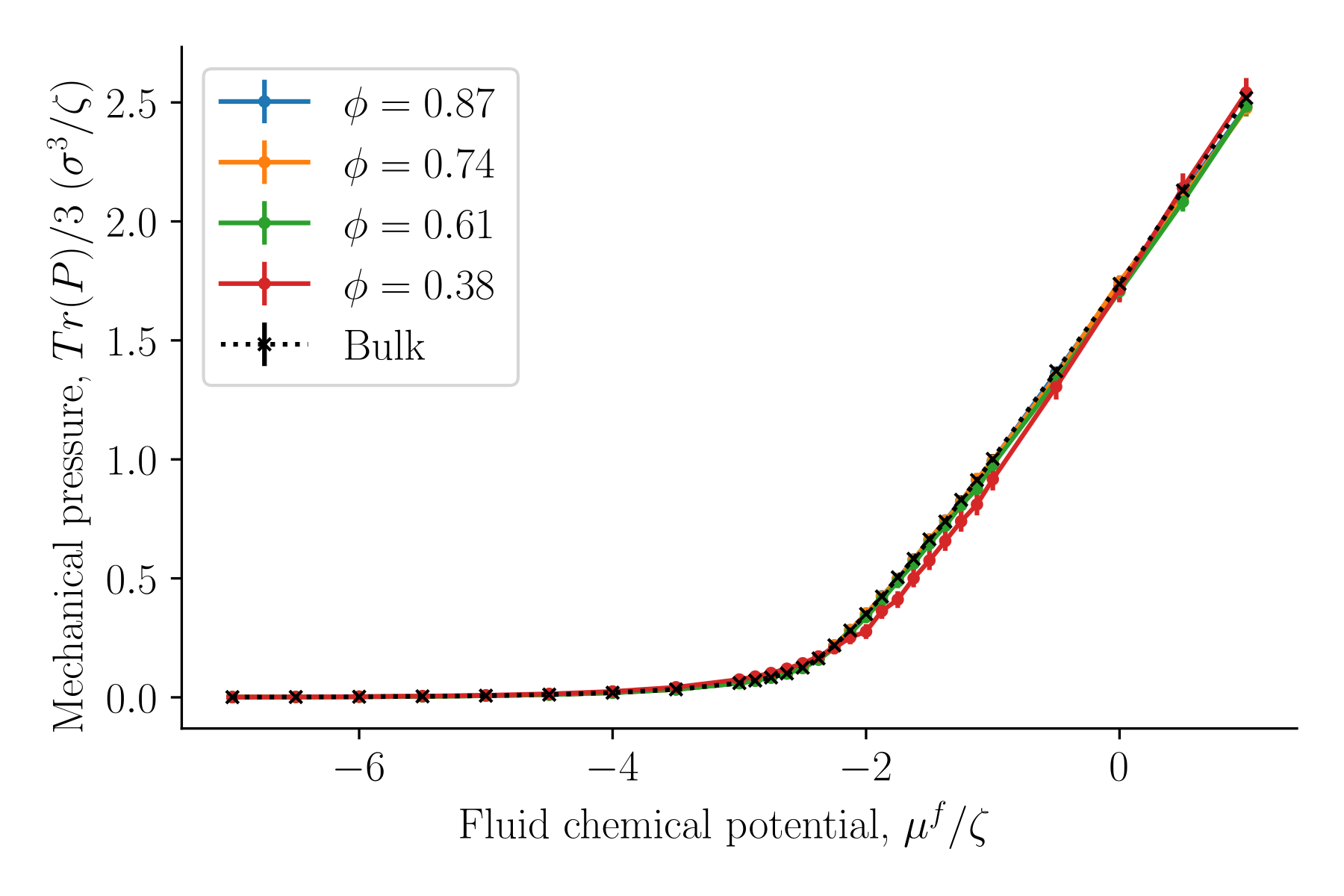}
    \caption{Trace of the mechanical pressure divided by three as a function of the fluid chemical potential $\mu^f$ for varying porosities $\phi=V^f/V$.}
    \label{fig:pressure}
\end{figure}

Figure \ref{fig:pressure} compares the average of the diagonal components of the mechanical pressure for the bulk fluid (black crosses) and the fcc lattice of spheres filled with fluid particles. The mechanical pressure was calculated in Cartesian coordinates for the whole simulation box, as described by equation \ref{eq:mechanical_pressure}. The bulk value was now equated to the thermodynamic bulk pressure, which was assumed to be equal to the integral pressure. Note that we did not equate the trace of the mechanical pressure tensor in the heterogeneous porous medium to any thermodynamic property. Figure \ref{fig:pressure} serves purely for a comparison. However, even though the geometry inside the porous medium is more complex than what can be captured by the Cartesian mechanical pressure tensor, the values fall almost on the same line. This could suggest that the mean of the diagonal components of the Cartesian mechanical pressure tensor gives the integral pressure. An exception is seen for the data obtained with the smallest porosity, $\phi=0.38$, when fluid chemical potentials vary between approximately $\mu=-2\zeta$ and $\mu=-\zeta$. The given porosity is near the closest packing possible for spheres. While this is interesting, it may only hold for the present case, with a relatively simple structure.  

\subsubsection{Internal energy and entropy densities}

\begin{figure}
    \centering
    \includegraphics[width=0.75\textwidth]{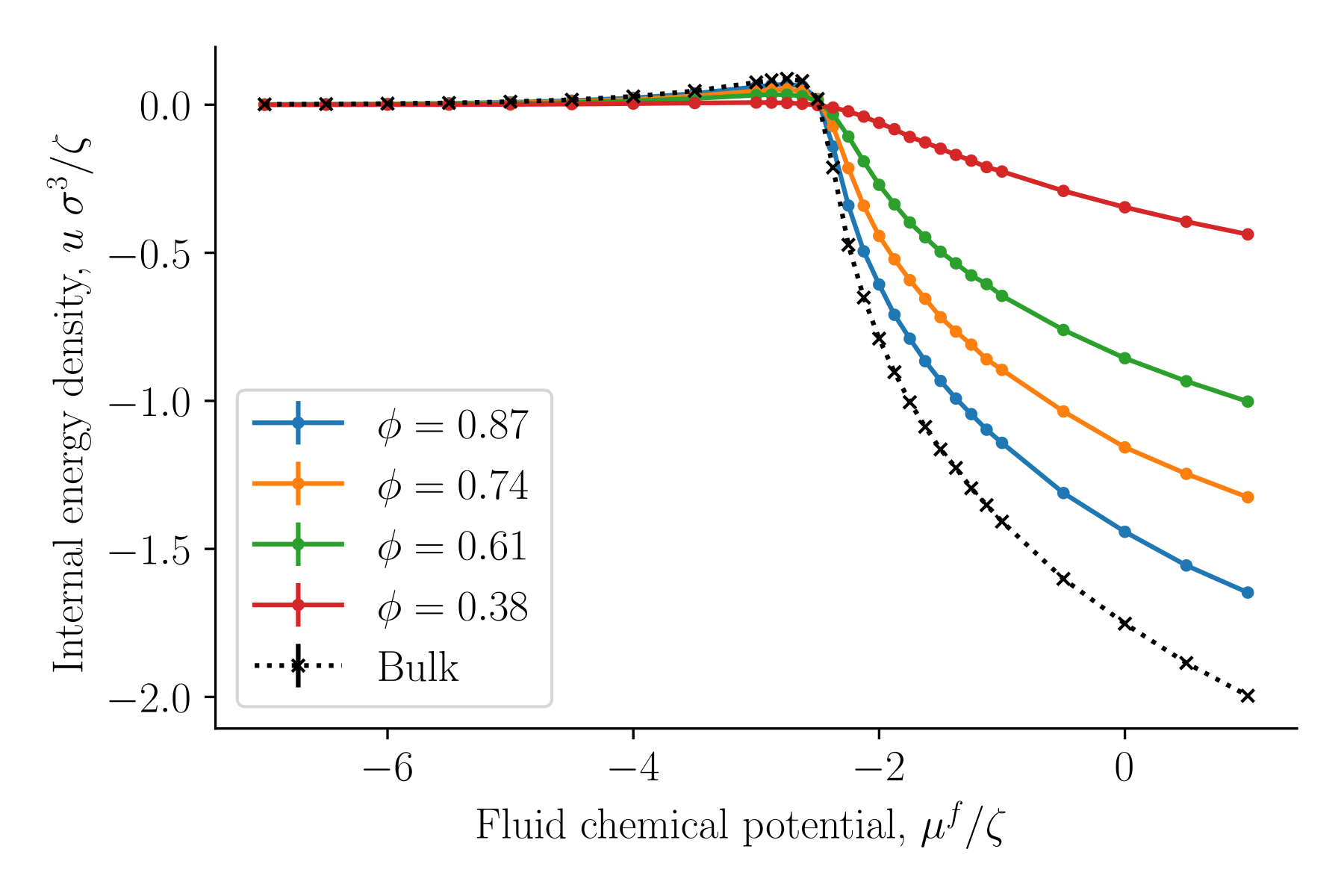}
    \caption{Internal energy density $u=U/V$ as a function of the fluid chemical potential $\mu^f$ for varying porosities $\phi=V^f/V$.}
    \label{fig:internal_energy_density}
\end{figure}

\begin{figure}
    \centering
    \includegraphics[width=0.75\textwidth]{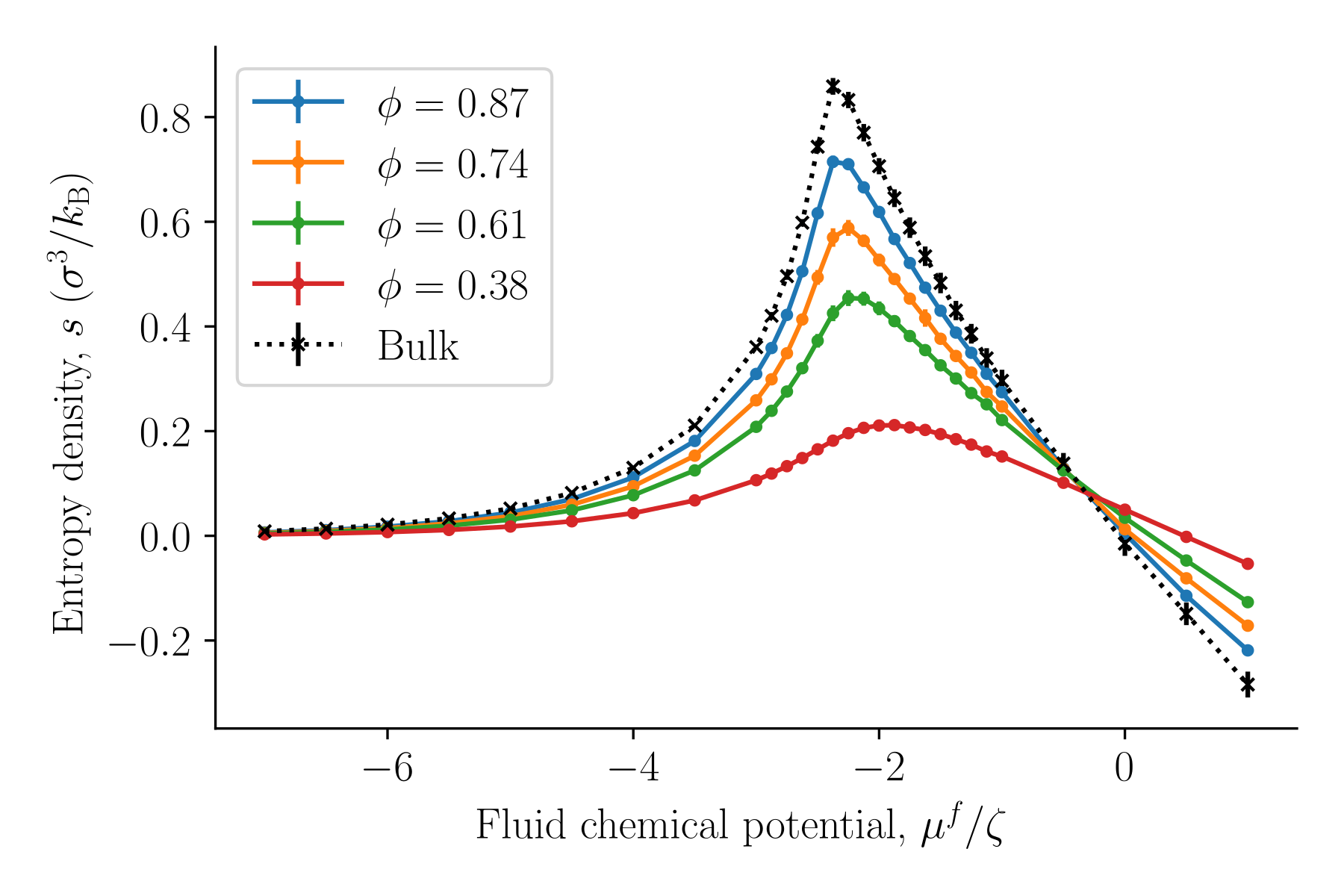}
    \caption{The entropy density $s=S/V$ as a function of the fluid chemical potential $\mu^f$ for varying porosities $\phi=V^f/V$. The integral pressure is assumed to be equal to the bulk pressure.}
    \label{fig:entropy_density}
\end{figure}

The internal energy and entropy densities as a function of the fluid chemical potential are presented in figure \ref{fig:internal_energy_density} and \ref{fig:entropy_density}, respectively. The internal energy and entropy are divided by the REV volume $V$ to give the respective densities. The variation in the internal energy and entropy densities as a function of the fluid chemical potential, is similar to that shown by fluids above the critical point, when they are described by cubic equations of state, for example, the van der Waals equations of state. It is therefore reasonable that the maxima of the internal energy and entropy densities are a consequence of a structural transition that takes place at the chemical potential in question. This will explain why all maxima are located at the same chemical potential (-2.3 $\zeta$). This location is close to the value of the chemical potential, where the fluid-solid surface tension starts to increase and also where the replica energy density begins to deviate stronger from its bulk value (see figures \ref{fig:surface_tension} and \ref{fig:replica_energy_density}). The maxima locations remain at the same chemical potential, as the porosity decreases and are given by the location in the bulk fluid. It may therefore be possible to estimate this position using the equation of state for the bulk fluid. Thus, one can determine for which fluid states it is particularly important to take the system size into account. 

The absolute value of the internal energy density decreased with the porosity, reflecting the fact that the fluid number density decreased with the porosity. The entropy density became larger as we approached the bulk value. This observation is typical for phase transitions in small systems, it becomes less clear cut first order, and has less of a discontinuity in the phase variables \cite{Hill1964}.   

\subsection{Small system effects}

A system is small when we need to take into account the subdivision potential to accurately describe the system. The smallness of a system can also be measured by the replica energy (for the present system, see equation \ref{eq:Replica}). The replica energy depends in general on the set of control variables, and has accordingly an equivalent bulk variable \cite{Bedeaux2020}. It changes proportional to the subdivision potential. The value of the subdivision potential is included in the deviation of the replica energy from the corresponding value of a bulk fluid, see equation \ref{eq:Replica}. This deviation can be seen in figure \ref{fig:replica_energy_density}; it increases with decreasing porosity. While a change in replica energy is expected with additional contributions of the solid, the application of the replica energy as a variable allow us to quantify the smallness of the system, the smallness that originate from the subdivision potential. For evaluation of the state of smallness, the single contributions given in equation \ref{eq:Replica} must be computed independent of one another. We were not able to compute the subdivision potential here, but the replica energy density may still be used to assess whether small system effects can be neglected or not. This can be done by evaluating the total replica energy difference together with the entropy density. Consider for the purpose of such an evaluation the left-hand side of the peak in entropy density. Due to the small replica energy differences between the fluid bulk value and the corresponding values of the nanoporous medium for small chemical potentials, it may be speculated that the subdivision potential is negligible for vapor-like densities. On the right hand side of the peak, the difference is sizeable which is why the subdivision potential cannot be assumed to be negligible a priori. It must therefore be included in the thermodynamic analysis to ensure first-order Euler homogeneity in the total internal energy. 

\subsection{The pressure of a REV in a porous medium} 

We are now in a position to discuss, if not answer completely, the questions posed upfront; what is the pressure of a representative elementary volume (REV)?

In the present case, the structure of the porous system was regular fcc, and all different microstates are represented by a small REV, a unit cell for all practical purposes. A system with irregular structure has probably a larger REV. All thermodynamic properties defined, refer to the actual REV.

Equations \ref{eq:dp_hat},\ref{eq:loce} and \ref{eq:phat} most central in the description. These relations were derived using Hill's systematic procedure for porous media with nanoscale pores. The effective pressure $\hat{p}$ of the REV was given in equation \ref{eq:dp_hat}. We see that the expression is consistent with Young-Laplace equation. Not only a set of system variables is central in the theory; also the set of control variables need be specified. When this is done, we can arrive, in the present case, at the expression linking $\hat{p}$ to two unknown system variables. 

In order to proceed, we have next taken the bold assumption stated in equation \ref{eq:loce}, that the integral pressure $\hat{p}$, which clearly differ from the differential pressure $p$, is constant across the medium at equilibrium. It is then not far from the next possible step; to construct the driving force for mass transport using $\hat{p}$.

The mean of the diagonal components of the Cartesian mechanical pressure tensor, presented in figure \ref{fig:pressure}, gives a good approximation of the integral pressure for the largest porosities. This can give an alternative route to obtain the integral pressure when the porosity is large. However, for small porosities it deviates from the bulk pressure and the integral pressure. 

\section{Conclusion}

The thermodynamic method of Hill for nanoscale systems was used to describe the thermodynamic state of a single-phase fluid confined to a porous medium.  The size and shape of the porous media were restricted, such that the description can be said to be general for any porous media. 

The system was open for fluid to be exchanged with the environment, while the solid was not allowed to be exchanged with the environment. In addition, the temperature was controlled by the environment, and the fluid and solid volumes, surface area, and number of solid particles were controlled. 

Nanothermodynamics introduces two new conjugate variables, the subdivision potential and the number of replicas. The subdivision potential was incorporated into the definition of the integral pressure and integral solid chemical potential. A fundamental assumption used in this work is that the integral pressure is constant everywhere in equilibrium. We have shown in previous works that this holds for simple porous media such as the slit pore \cite{Rauter2020,Galteland2021}. 

We have used this framework to demonstrate how to compute the fluid number density, replica energy density, integral solid chemical potential, internal energy density, and entropy density of a fluid confined to a face-centered cubic lattice of solid particles. The radius of the solid particles was ten times larger than the fluid particles, and the porosity varied from $\phi=0.38$ to $\phi=0.87$. These porosities range from almost closest packing to rather open structures. We have used this system as a relatively simple model of porous media, computing all its thermodynamic properties, in particular its integral and differential pressures, liquid-solid surface tension, and solid pressure of a single solid particle surrounded by fluid particles. 

This way of defining the small system thermodynamic properties, for a given set of control variables, may be useful for the study of transport in non-deformable porous media. 

\section*{Conflict of Interest Statement}
The authors declare that the research was conducted in the absence of any commercial or financial relationships that could be construed as a potential conflict of interest.

\section*{Author Contributions}
O.G. contributed to formal analysis, investigation, methodology, software, and visualization. D.B. and S.K contributed to supervision. All authors contributed to conceptualization, writing original drafts, reviewing, and editing. All authors have read and agreed to the published version of the manuscript.

\section*{Funding}
This work was funded by the Research Council of Norway through its Centres of the Excellence funding scheme, project number 262644, PoreLab.

\section*{Acknowledgments}
The simulations were performed on resources provided by UNINETT Sigma2 - the National Infrastructure for High-Performance Computing and Data Storage in Norway. We thank the Research Council of Norway.

\bibliographystyle{ieeetr}
\bibliography{arxiv_library}
\end{document}